\documentclass[12pt]{article}
\usepackage{graphicx}
\usepackage{nico}

\newcommand\pubnumber{}
\newcommand\pubdate{\today}

\def\edinburgh{School of Physics and Astronomy, University of Edinburgh,\\ Edinburgh EH9 3JZ, U.K.}

\def\Title#1{\begin{center} {\Large #1 } \end{center}}
\def\Author#1{\begin{center}{ \sc #1} \end{center}}
\def\Address#1{\begin{center}{ \it #1} \end{center}}

\newcommand\pubblock{\rightline{\begin{tabular}{l} \pubnumber\\
         \pubdate  \end{tabular}}}
\newenvironment{Abstract}{\begin{quotation}  }{\end{quotation}}
\newenvironment{Presented}{\begin{quotation} 
      \begin{center}\begin{large}}{\end{large}\end{center} \end{quotation}}

\begin{document}
\begin{titlepage}
\pubblock

\vfill
\Title{Lattice determination of $\fBd$, $\fBs$, and $\xi$}
\vfill
\Author{ Nicolas Garron}
\Address{\edinburgh}
\vfill
\begin{Abstract}
The search for new physics is very often limited by the size of the theoretical 
uncertainties, mainly due to the errors affecting the hadronic quantities. 
This is especially true in the $B$-physics area, where in many cases the errors
are largely dominated by quantities like 
heavy-light decay constants or bag parameters,
which are obtained by lattice simulations. 
If the recent progress of the lattice community in the light quark sector 
is very impressive, lattice simulations around the $b$-quark mass are still difficult
(recent lattice progress has been reviewed at this conference in~\cite{Junko:ckm2010}).
In this talk, I summarize the recent lattice determinations of the decay constants and of the bag parameters
of the heavy-light and heavy-strange neutral mesons. 
In the next section, I remind the reader where lattice computations enter in neutral B meson phenomenology.
In section~2, I briefly describe the different lattice techniques 
for the quark b, and in section~3, I present and comment some recent lattice results,
and discuss the main advantages and the disadvantages of the various methods used to simulate the quarks.

\end{Abstract}
\vfill
\bigskip
\begin{Presented}
Proceedings of CKM2010, the 6th International Workshop on the CKM Unitarity Triangle, University of Warwick, UK, 6-10 September 2010
\end{Presented}
\vfill
\end{titlepage}
\def\thefootnote{\fnsymbol{footnote}}
\setcounter{footnote}{0}
%


\section{Reminders}

\begin{figure}[t]
\centering
\includegraphics[height=3.cm]{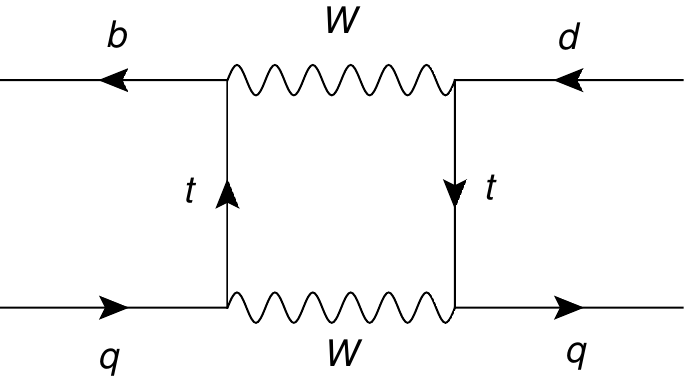}
\caption{Example of a box diagram which dominates $B_{(\rm q)}-\bar B_{(\rm q)}$ mixing ($\rm q=d,s$).}
\label{fig:BBmix}
\end{figure}

The physical eigenstate $\Bq^{L,H}$ ($\rm q\in\{d,s\}$) are related to the flavor eigenstate
$\Bq, \Bqbar$ by
(see for example~\cite{Nakamura:2010zzi})
\be
\Bq^{L,H}=\alpha\Bq \pm \beta \Bqbar \;,\qquad |\alpha|^2+|\beta|^2=1 \;.
\ee
Experimentally, by measuring the frequencies of the B oscillations, 
one can access to the differences of mass 
\be
\Delta \mq= m_{\Bq^{H}}-m_{\Bq^{L}} \;.
\ee
In the standard model this quantity is dominated by box diagrams
with t-quark and W exchanges (see fig~\ref{fig:BBmix}).
After performing an OPE to separate the long distance physics from the short distance one, 
one finds
\be
\label{eq:Dmq}
\Delta \mq = C \mBq \fBq^2 \BBq |\Vtq\Vtbs| \;
\ee
where C is a (known) numerical constant which contains the short distance effects, 
and $V_{ij}$ are CKM matrix elements.  
The non-perturbative part is factorized in the mass $m_{\Bq}$, 
the decay constant $f_{\Bq}$ and the bag parameter 
$\BBq$ of the heavy-light neutral meson $\Bq$ :
\bea 
\la 0 | A_0| \Bq \ra &=&  \fBq \mBq \;,\\
{
\la \Bqbar | (\bar b \gamma_\mu^L q) (\bar b \gamma_\mu^L q) | \Bq \ra }
&=& 
{8 \over 3} {\BBq} \mBq^2 \fBq^2 \;.
\eea
The unitarity of the CKM matrix implies twelve distinct complex relations 
among the matrix elements, including
\be
\Vud \Vubs + \Vcd\Vcbs +\Vtd\Vtbs=0 \;,
\ee
which is traditionally represented by a triangle in the complex plane
$(\bar \rho, \bar \eta)$, where $\bar \rho$ and $\bar \eta$ 
are Wolfenstein parameters. 
Hence, setting $q=d$ in eq.(\ref{eq:Dmq})
one sees that 
a computation of $\fBd^2 \BBd$ implies a constraint on $\Vtd\Vtbs$
(so on the length of one side of the triangle). 
In the next section I will discuss some difficulties of 
such a computation.
Because it is expected that some uncertainties cancel in the ratio 
of bag parameters, one can 
compute on the lattice the 
$SU(3)$ breaking ratio
$
\xi={\BBs\fBs^2 \over \BBd\fBd^2} \;,
$
and then use the experimental value of $ \Delta \ms$ in the relation
$
{\Delta \ms \over \Delta \md} = {\mBs \over \mBd} \; \xi^2 \left| {\Vts \over \Vtd} \right|^2 \;.
$
Finally we note that the decay constants and bag parameters of the B mesons also appear in other quantities,
like the decay rate differences $\Delta_\Gamma$ and the semileptonic CP asymmetries,
which also play an important role in the search for new physics
(see eg~\cite{Lenz:2010gu}).

\section{Lattice computation}

Lattice QCD provides a natural framework 
for the computation of heavy-light decay constant 
and bag parameters. 
However, such computations are not that easy for different reasons.
First 
one has to compute the matrix element of an 
operator (a four quark-operator for the bag parameters)
and this matrix element has to be renormalized, non-perturbatively if possible.
This renormalization can be complicated if the discretization of the 
quarks breaks chiral symmetry, which is the case for example 
if Wilson-type fermions are employed. 
But the main difficulty comes from the heavy-light nature
of the particle considered.
On the one hand, in order to keep the discretization effects under control 
when one simulates a heavy quark of mass $\mQ$,
one needs a fine lattice spacing $a$, typically $a\ll \mQ^{-1} \sim 0.04 \fm $ for a b quark.
On the other hand the simulated volume should be large 
enough 
to contain the light degrees of freedom,
for example one would like the space extent $L$ of the lattice to be larger
than the Compton wavelength of a pion $L\gg m_\pi$. 
Putting these two constraints together, one see 
that a very large number of points $(L/a)^4$ is needed. \\

Various strategies have been developed 
in the literature to circumvent this problem
(see e.g.~\cite{Aubin:2009yh, DellaMorte:2010ym} 
for recent lattice review).
First, one can 
use an effective theory 
for the b-quark, like non-relativistic QCD (NRQCD)~\cite{Thacker:1990bm},
or heavy quark effective theory (HQET)~\cite{Eichten:1989zv}.
In that case one replaces the QCD Lagrangian in the heavy sector 
by an expansion in the inverse quark mass 
and in the velocity.
For the heavy-light systems, 
one difference between these effective theories 
is the way 
$1/\mQ$ terms (or higher order), like for example the kinetic term (${D_{\perp}}^2/(2m)$, 
are treated.
In HQET, only the leading order (where the $b$ quark 
is infinitely heavy) is kept in the exponent of the path integral, 
and corrections in inverse power of the heavy quark mass 
are inserted into the static Green function.
In NRQCD, all the terms (up to a given order in the heavy quark expansion) 
are kept in the Dirac operator.
An important consequence (see e.g.\cite{Sommer:2010ic}) is that the continuum 
limit $a\to 0$ cannot be taken in NRQCD, in contrast 
with HQET where the continuum limit exists
if the theory is non-perturbatively renormalized. \\

Another way to simulate heavy quark on the lattice is given by the 
Fermilab formulation~\cite{El-Khadra:1996mp},
where one uses on-shell Symanzik improvement, treating both $a$ and $1/\mQ$ as short
distances, in such a way that the theory still makes sense when $a\mQ>1$. 
By construction the theory reproduces an effective theory in the heavy quark region
by a matching to HQET or NRQCD, and is still well-defined in the limit $a\to0$.
Probably inspired by the Fermilab action, other relativistic heavy quark 
formulations have been proposed in the literature~\cite{Aoki:2001ra,Christ:2006us}
where one advantage of the latter is the fact that the matching can be done
non-perturbatively~\cite{Lin:2006ur} (a study of the heavy-light and heavy-strange
decay constant along this line has been presented very recently~\cite{VandeWater:2011gr}). 
\\

It is also possible to use the effective theory as a guide, 
as done for example, in~\cite{DellaMorte:2007ij} 
by the Alpha collaboration in a (quenched) computation of the decay constant $\fBs$.
They first compute a decay constant at several  ``heavy'' quark masses
around the charm, then they redo the computation with 
an infinitely heavy (static) $b$ quark, and they interpolate 
to the mass of the $b$ to obtain $\fBs$. 
Finally let us mention a proposal by ETMC~\cite{Blossier:2009hg},
which is somehow a variant of the previous method.
The authors construct ratios of quantities possessing a known static limit,
evaluate them at various pairs of heavy quark masses around the charm,
and interpolate to the physical point.

\section{Recent Lattice results}

The lattice simulations are now systematically taking into account the sea quark effects
(or at least the dominant ones).
Here I discuss some results obtained with $n_f=2$ or $n_f=2+1$ 
(meaning a doublet of degenerate quarks for the $u$ and the $d$ and 
a heavier quark for the $s$) flavors of sea quarks. 
Various discretizations of the light quarks (both in the valence and in the sea sectors) 
are available on the market. Because also the light quark masses are expensive,
a chiral extrapolation is in general necessary to reach the physical light quark masses.
In table~\ref{tab:results}, I summarize some recent results for the decay constants
and for the bag parameters, together with the chosen discretization (or the strategy)
for the heavy and for the light quark, and the number of dynamical flavors.
In the following I comment briefly the different aspects 
of the various formulations, starting by the $n_f=2+1$ flavors case.
\\

The FNAL/MILC collaborations~\cite{Bernard:2009wr,Evans:2009du}
employ the Fermilab formulation 
for the heavy quark, on the $n_f=2+1$ MILC ensembles.
They use three values of the lattice spacing $a\sim 0.09, 0.12, 0.15 \fm$ 
in order to extrapolate to the continuum limit. 
The light quarks are described by the improved-staggered fermions called Asqtad.
Such fermions are numerically cheap and exhibit good chiral properties, 
but because they use a ``rooting'' procedure, one can argue that they 
might not fall into the right universality class.
A discussion whether or not one should employ such a formulation in a lattice simulation 
is beyond the scope of this review. \\

HPQCD~\cite{Gamiz:2009ku} uses the same MILC ensembles 
(and then also the same discretization of the light quarks)
and NRQCD for the heavy quark, implying as discussed earlier that the continuum limit cannot 
be taken. Nevertheless, two different values of the lattice spacing $a\sim 0.09, 0.012\fm$ 
have been studied, and the authors fit their result to a functional form, performing then 
together the chiral and the continuum extrapolations.\\ 

The RBC-UKQCD\cite{Albertus:2010nm} collaborations employ a completely different approach. 
The light quark masses are simulated with a Domain-Wall action 
(which possess an almost exact chiral-flavor symmetry), 
and the heavy quark mass is taken in the static limit. This is a theoretically interesting
framework, since the continuum limit can be taken, the discretization effects 
are expected to be small and the renormalization is simplified by the good 
chiral properties of the light fermions.
The main drawback of this approach is probably the numerical cost, 
because the static action is noisy by nature, and the Domain-Wall fermion 
numerically expensive, it is hard to obtain a good signal-over-noise ratio. 
For this reason, only one lattice spacing was simulated. 
\\

We turn now to the $n_f=2$ simulations. 
ETMC has computed the decay constants $\fBd$ and $\fBs$ (but no bag parameters)
using twisted-mass fermions. In this formulation, one can obtain automatic O(a) 
improvement (ie no order $a$ discretization effects) without having to pay 
the price of a chiral formulation.
For example in the case of four quark operators, one obtains the same 
simple renormalization pattern as in the continuum.
The disadvantages of this formulation is that one breaks isospin symmetry, and that only $n_f=2$
(or $n_f=2+1+1$) flavors or dynamical quarks can be simulated.  
The two methods employed for the heavy quark rely on an interpolation 
between the static approximation and the charm region. 
Depending on the method, three or four different lattice spacings 
varying in the range $0.064 \fm < a < 0.1 \fm $ have been simulated. \\

Finally the Alpha collaboration uses non-perturbative HQET, 
developed at the $\Lambda_{\rm QCD}/\mb$ order.
The key idea is to match HQET to QCD in a small volume in order to obtain
the HQET parameters. This is theoretically advantageous, because the power divergences
cancel explicitly and thus the continuum limit can be taken. 
Moreover one expects the remaining $(\Lambda_{\rm QCD}/\mb)^2$ effects to be small. 
The light quarks are described by O(a)-improved fermions.
Even if only a preliminary result of $\fB$ is available~\cite{Blossier:2010vj}, 
where a single (but rather fine) lattice spacing $a\sim0.07\fm$ has been considered so far,
this is a very encouraging direction for the future.\\

For the decay constants, one notices that all the results
are in good agreement, despite the different formulation used. 
Concerning the SU(3)-breaking ratio, 
the only non-staggered result has been obtained by RBC-UKQCD,
and even if the quoted error is larger than the ones given by 
FNAL/MILC and by HPQCD (which are based on the same MILC ensembles),
we see that also here all the results are compatible. 

\begin{table}[t]
\begin{center}
\begin{tabular}{ccccccc}
Group  &    $ \fBd  $        & $ \fBs$ & $ \xi$  &  $ n_f$  & Heavy  & Light  \\
       & $(\MeV)$                 & $(\MeV)$  &         &         &        & 
\\
\hline
\\
{ FNAL/MILC} & $ 195(11)^* $   & $ 243(11)^*$  & $ 1.205(50)^*$ & $ 2+1$   & Fermilab  &  Asqtad \\
 HPQCD    & $ 190(13) $   & $ 231(15)$  & $ 1.258(33)$ & $ 2+1$   & NRQCD     &  Asqtad \\
 RBC-UKQCD &                      &            & $ 1.13(12)$  & $ 2+1$    & Stat.    & DW \\ 
 ETMC      & $ 191(14)^* $   & $ 243(13)^*$         &            & $ 2$      & Stat. +Int. &  TM \\
 ETMC      & $ 194(16) $   & $ 235(12)$         &            & $ 2$      & Stat. +Int.  &  TM \\
ALPHA      & $ 178(16)^* $ &                     &             & 2         & Stat+$1/m$  & Clover \\
\end{tabular}
\caption{Selection of recent lattice results. I took the liberty to add in quadrature 
the statistical and the systematic errors.
$^*$ refers to results which have been published only in proceedings
(and the error budget might be incomplete). 
``Stat.'' stands for static, ``Int.'' for interpolation, ``DW'' for Domain-Wall and ``TM'' for Twisted Mass.}
\label{tab:results}
\end{center}
\end{table}

\section{Conclusions and outlook}
Lattice QCD is making very good progress, even in the heavy quark sector 
it becomes now possible to obtain precise results. 
But there is still a lot of room for improvement,  
for example, concerning the bag parameters, the results are largely  
dominated by formulations which employ rooted staggered fermions 
in the light sector.
It is important that the lattice community gives also precise results 
with other light quark formulations.
Different approaches to treat the heavy quark on the lattice,
in particular with more solid theoretical foundations, have been developed and tested 
in the last years. 
I believe that such approaches, where one gets a much better handle on the systematics errors, 
are fundamental to constrain the standard model, and hopefully see the effects of new physics. 
Moreover, we also need a better control of the chiral extrapolations 
(so pion masses closer to the physical one) 
and of the discretization effects (eg by using three or more lattice spacings on a range $\sim 0.05\fm -0.1 \fm$).
If the charm quark is quenched, this effect has to be quantified as well. 
Finally let us mention that some quantities like beyond the standard model contributions 
to B mesons mixing are still missing a computation with dynamical 
fermions~\footnote{Although at least one computation is on the way~\cite{Bouchard:2010yj}.} .

I would like to thank the organizers for inviting me to give this talk,
and for a very enjoyable conference.

\bibliography{lattB}{}
\bibliographystyle{h-elsevier}

\end{document}